\documentclass[conference]{IEEEtran}
\IEEEoverridecommandlockouts
\usepackage{cite}
\usepackage{amsmath,amssymb,amsfonts}
\usepackage{algorithmic}
\usepackage{graphicx}
\usepackage{textcomp}
\usepackage{xcolor}
\def\BibTeX{{\rm B\kern-.05em{\sc i\kern-.025em b}\kern-.08em
    T\kern-.1667em\lower.7ex\hbox{E}\kern-.125emX}}

\usepackage[ruled,vlined]{algorithm2e}
\usepackage{xurl}

\usepackage{pgfplots}
\pgfplotsset{compat=1.16}

\usepackage{tikz}

\usepackage{hyperref}

\usepackage{booktabs}
\usepackage{tabularx}
\usepackage{array}

\usepackage{makecell}

\newcolumntype{R}{>{\raggedleft\arraybackslash}X}

\renewcommand{\thetable}{\arabic{table}}

\renewcommand{\figurename}{Figure}
\renewcommand{\tablename}{Table}

\begin{document}

\title{
FeatherWallet: A Lightweight Mobile Cryptocurrency Wallet Using zk-SNARKs
}

\author{
	\IEEEauthorblockN{Martin Pere\v{s}\'{i}ni, \href{mailto:iperesini@fit.vut.cz}{iperesini@fit.vut.cz}}	
        \IEEEauthorblockN{Ivan Homoliak, \href{mailto:homoliak@vut.cz}{homoliak@vut.cz}}
        \IEEEauthorblockN{Samuel Olek\v{s}\'{a}k,
    \href{mailto:ioleksak@fit.vut.cz}{ioleksak@fit.vut.cz}}
	\IEEEauthorblockN{Samuel Sl\'{a}vka, \href{mailto:slavka.samuel@gmail.com}{slavka.samuel@gmail.com}}	
    
    \IEEEauthorblockA{\textit{Faculty of Information Technology}, \textit{Brno University of Technology}\\
		Czech Republic}
}

\maketitle

\begin{abstract}
Traditionally, mobile wallets rely on a trusted server that provides them with a current view of the blockchain, and thus, these wallets do not need to validate the header chain or transaction inclusion themselves.
If a mobile wallet were to validate a header chain and inclusion of its transactions, it would require significant storage and performance overhead, which is challenging and expensive to ensure on resource-limited devices, such as smartphones.
Moreover, such an overhead would be multiplied by the number of cryptocurrencies the user holds in a wallet.

Therefore, we introduce a novel approach, called FeatherWallet, to mobile wallet synchronization designed to eliminate trust in a server while providing efficient utilization of resources.
Our approach addresses the challenges associated with storage and bandwidth requirements by off-chaining validation of header chains using SNARK-based proofs of chain extension, which are verified by a smart contract.
This offers us a means of storing checkpoints in header chains of multiple blockchains.

The key feature of our approach is the ability of mobile clients to update their partial local header chains using checkpoints derived from the proof verification results stored in the smart contract. In the evaluation, we created zk-SNARK proofs for the 2, 4, 8, 16, 32, and 64 headers within our trustless off-chain service.
For 64-header proofs, the off-chain service producing proofs requires at least 40\,GB of RAM, while the minimal gas consumption is achieved for 12 proofs bundled in a single transaction.
We achieved a 20-fold reduction in storage overhead for a mobile client in contrast to traditional SPV clients.
Although we have developed a proof-of-concept for PoW blockchains, the whole approach can be extended in principle to other consensus mechanisms, e.g., PoS.

\end{abstract}

\IEEEpeerreviewmaketitle

\section{Introduction}
\label{sections:intro}

Mobile wallets typically rely on trusted centralized services to provide synchronization with the blockchain.
However, there have been many incidents in which trusted centralized services and exchanges were compromised, leading to significant financial losses for users.
Notable examples include the Mt. Gox hack in 2014~\cite{2014-Mt-Gox}, where approximately 850,000 bitcoins were stolen, the Bitfinex hack in 2016~\cite{2016-Bitfinex-hack} which led to the loss of \$72 million, and the Solana mobile wallet exploit, resulting in the theft of over \$8 million in SOL~\cite{securityweek2023crypto, chainalysis2023security}. 
The Coincheck hot wallet breach~\cite{coincheck}, which resulted in a \$534 million hack.

Another SPV wallet-specific vulnerability is the \textit{false-negative}~\cite{decred_spv} attack, where a malicious peer sends a fake filter to the wallet, preventing it from downloading the relevant block and becoming aware of a transaction.
Additionally, vulnerabilities in Merkle tree construction can be exploited for \textit{Fake Transaction Attacks} (FTAs)~\cite{fta}, allowing an attacker to simulate payments that never actually happened.

In 2024 alone, \$2.36 billion~\cite{2024-CertiK-report} in cryptocurrencies was lost within 760 security breaches, highlighting the ongoing risks associated with centralized trusted platforms.

The decentralized nature of blockchain technology requires that the full nodes of the blockchain store the entire chain locally. However, as the size of the blockchain grows linearly, maintaining local copies of the entire blockchain becomes expensive, particularly for resource-constrained devices.
The introduction of lightweight clients~\cite{light-clients}, intended for such devices, allowed the reduction of the network and storage overhead by storing only a part of the blockchain -- its headers.

Despite the advancements of lightweight clients and the increasing storage space, the continuous growth of the header chain has made their storage unsustainable~\cite{plumo}, especially in the case of multiple blockchains synchronized by a single client.
This increasing demand for more lightweight solutions has led to a shift toward ultralight clients~\cite{plumo} or hosted wallets in the mobile client ecosystem such as Blockchain Wallet~\cite{BlockchainInfoWallet}, Coinbase~\cite{CoinbaseWallet} or Binance~\cite{binance-exchange}, which rely on a centralized party.
In another direction, Vesely et al. introduced Plumo~\cite{plumo}, an ultralight client that uses SNARKs to minimize resource usage while maintaining security for light clients without trust in any centralized party.
However, Plumo supports only the Celo blockchain~\cite{celo}. 

\textbf{\textit{Proposed Approach.}}
Addressing the growing storage and computational demands of blockchain clients is crucially important.
In response to these challenges, we propose an optimized approach for mobile cryptocurrency wallets that supports multiple proof-of-work blockchains.
Our solution integrates zk-SNARKs within a client-server architecture, where servers 
generate proofs for sequences of blocks and submit them to the blockchain for on-chain verification.
This process establishes trust in the integrity of the blockchain for the client.

\label{sections:contrib}
\textbf{\textit{Contributions.}}
Our contributions are as follows:
\begin{itemize}
    \item
    We proposed a client-server approach that integrates zk-SNARKs to optimize the storage and network overhead of mobile wallets that support multiple PoW-based blockchains.
    \item We made a proof-of-concept of the proposed approach, consisting of a client (mobile application) and a server.
    \item We evaluated our approach and demonstrated its efficiency in terms of gas consumption and performance.
\end{itemize}

\section{Background}
\label{sections:background}
\subsection{ZKPs and Their Application in Blockchains}

Zero-knowledge proofs (ZKPs) are cryptographic protocols that allow parties to verify the correctness of computations~\cite{howsnarks, zkp1} without revealing any additional information beyond the truth of the statement being proven.
This property is particularly valuable in blockchain technologies, where privacy and security are paramount.
There are multiple ZKP use cases in blockchains:
\textbf{\textit{(1) Cross-chain Interoperability}}:
ZKPs enable seamless communication between different blockchain networks by allowing one network to verify transactions or state changes on another without exposing sensitive data~\cite{zk-relay}.
\textbf{\textit{(2) Private Transactions}}:
Using ZKPs, the sender and receiver can confirm the validity of the transaction without revealing any personal or transaction-specific information~\cite{tornado-cash}.
\textbf{\textit{(3) Scalable and Secure Layer-2 Facilitation}}:
ZKRollups~\cite{zkrollup, zksync}, a type of Layer-2 solution, use ZKPs to aggregate multiple transactions into a single proof.
This reduces the load on the main chain, improving scalability while maintaining security.
\textbf{\textit{(4) Lightweight Blockchains}}:
Recursive ZKPs, as used in the Mina~\cite{mina_protocol} protocol, allow the creation of lightweight blockchains.
These blockchains can maintain a full state without storing the entire transaction history, thus reducing storage requirements and improving efficiency.
\textbf{\textit{(5) Decentralized Identity and Authentication}}:
ZKPs can be used to verify identities and authenticate users~\cite{zk_login} without revealing unnecessary personal information.

\begin{table}[t]
    \centering
    \footnotesize
    \renewcommand{\arraystretch}{1.5}
    \setlength{\arrayrulewidth}{0.5pt}
    \caption{Comparison of various zero-knowledge protocols.}
    \label{tab:zkcomparison}
    \begin{tabularx}{\linewidth}{>{\bfseries}l| *{3}{X}}
        \toprule[1.5pt]
        \textbf{Feature} & \textbf{zk-SNARKs} & \textbf{zk-STARKs} & \textbf{Bulletproofs} \\
        \midrule
        
        \textit{Cryptography} & 
        Elliptic curves & 
        Hashing & 
        Elliptic curves \\
        
        \textit{Trusted Setup} & 
        Required & 
        \makecell[l]{Not Required\\ (Transparent)} & 
        \makecell[l]{Not Required\\ (Transparent)} \\
        
        \textit{Proof Size} & 
        Very Small & 
        Larger & 
        Small \\
        
        \textit{Verification Speed} & 
        Fast & 
        Slower & 
        Slower \\
        
        \textit{Proof Generation} & 
        \makecell[l]{Slower for\\ complex probl.} & 
        \makecell[l]{Scalable, fast} & 
        \makecell[l]{Slower for\\ complex probl.} \\
        
        \textit{Scalability} & 
        \makecell[l]{Limited} & 
        \makecell[l]{Designed for\\ scalability} & 
        \makecell[l]{Intended for\\ smaller proofs} \\
        \bottomrule[1.5pt]
    \end{tabularx}
    \vspace{-1.2em}
\end{table}

\subsection{Zero-Knowledge Proofs}
Among the various types of ZKPs, zk-SNARKs~\cite{zksnarks1} and zk-STARKs~\cite{zkstarks} are prominent, each offering distinct trade-offs in terms of efficiency, security assumptions, and implementation complexity.
Bulletproofs~\cite{bulletproofs} represent another category of ZKPs, providing alternative characteristics.

\textbf{\textit{zk-SNARKs.}}
Zero-Knowledge Succinct Non-interactive Arguments of Knowledge are a type of ZKP characterized by their succinct proof size and non-interactive nature.
Succinctness projects into the proof size of only a few hundred bytes, regardless of the complexity of the statement being proven.
Non-interactiveness implies that once the prover generates the proof, the verifier can check its validity without further interaction with the prover.
A limitation for many zk-SNARK applications is the requirement for a trusted setup.

\textbf{\textit{zk-STARKs.}}
Zero-Knowledge Scalable Transparent Arguments of Knowledge emerged as an alternative to zk-SNARKs, addressing some of their limitations, particularly the trusted setup.
The key innovation of zk-STARKs is their transparency, which means that they do not require a trusted setup.
The common parameters used are publicly verifiable and can be generated using publicly known randomness, eliminating the risk of malicious setup.
zk-STARKs typically generate proofs that are larger than zk-SNARK proofs, and require longer verification times~\cite{zkbench}.

\textbf{\textit{Bulletproofs.}}
Bulletproofs are known for the elimination of a trusted setup alike zk-STARKs while offering relatively short proofs (although larger than zk-SNARKs proofs but smaller than zk-STARKs proofs).
\autoref{tab:zkcomparison} shows a comparison of different zero-knowledge protocols.

\subsection*{Backends for Zero-Knowledge Proofs}
To implement a zero-knowledge proof system, one needs to select a specific cryptographic backend.
The backend defines the concrete algorithms for proving and verifying, the underlying mathematical structures, and ultimately impacts the efficiency and security of the ZKP scheme.
Groth16~\cite{groth16}, PLONK~\cite{plonk}, and FFLONK~\cite{fflonk} are examples of such backends, specifically for zk-SNARKs.

\begin{figure}[t]
	\centering
	\includegraphics[width=\linewidth]{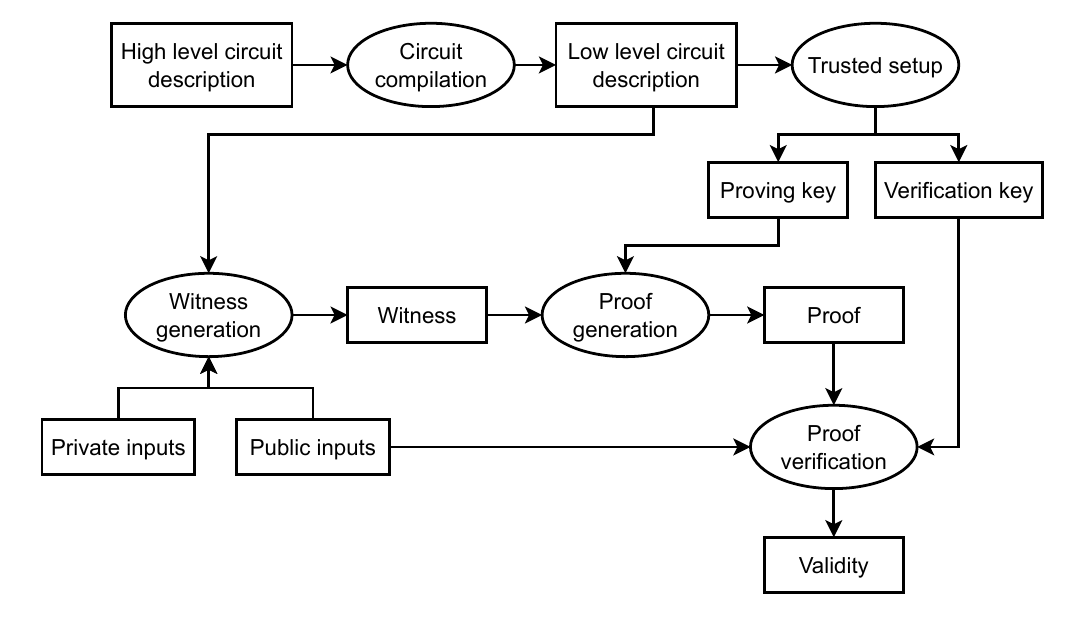}
	\caption{Individual steps of zero-knowledge proof generation and verification.}
	\label{fig:zk-phases}
    \vspace{-1.2em}
\end{figure}

\subsection{Zero-Knowledge Ecosystem}
Implementing ZKPs can be complex due to intricate cryptographic and mathematical operations.
To simplify this process, several software frameworks have emerged that abstract away much of the underlying complexity and allow developers to define and use ZKPs more efficiently.
These frameworks generally follow a structured workflow, as depicted in \autoref{fig:zk-phases}, which can be broadly divided into distinct phases:

\textbf{High-Level Circuit Description}: Define the computation or statement that needs to be proven in zero-knowledge using a high-level domain-specific language (DSL).

\textbf{Circuit Compilation}: Compile the high-level description into an arithmetic circuit, translating program logic into polynomial equations over a finite field.
This phase is computationally intensive for complex programs. 

\textbf{Low-Level Circuit Description}: Represent the computation as a series of arithmetic gates and constraints suitable for the chosen ZKP backend.

\textbf{Trusted Setup}: Generate common reference parameters, which are exported as proof and verification keys.
Secure multiparty computation (MPC) protocols are often used to mitigate risks.
However, as mentioned above, some frameworks (zk-STARKs, Bulletproofs) eliminate a trusted setup.
This phase is usually computationally intensive to ensure secure randomness.

\textbf{Witness Generation}: Compute intermediate values and satisfy all constraints based on public and private inputs.
A valid witness ensures the proof's validity.

\textbf{Proof Generation}: Use the valid witness and proving key to generate a compact zero-knowledge proof.
This phase is usually computationally intensive. 

\textbf{Proof Verification}: Check the proof's validity using the verification key and public inputs.
Proof verification is generally much faster than proof generation.
The succinctness property of ZKPs ensures that the verification time is typically shorter than that of the original statement.

Several tools and frameworks have been developed, such as ZoKrates~\cite{zokrates}, circom/snarkjs~\cite{circom, snarkjs}, gnark~\cite{gnark} and libsnark~\cite{libsnark}, to facilitate the implementation of ZKPs.

\section{Proposed Approach}
\label{sections:approach}
Our proposed approach addresses efficient blockchain verification for resource-constrained mobile wallets using an off-chain mechanism with zero-knowledge proofs.
This reduces synchronization overhead by minimizing the data clients need to process, overcoming storage and bandwidth limitations.
Traditional mobile wallets struggle with verifying their transactions due to downloading and validating many block headers, which consume considerable storage and bandwidth.
This is further exacerbated in the case of maintaining wallets to multiple blockchains.
Our solution enables lightweight verification without full blockchain synchronization, making it practical for mobile devices.
At the same time, our solution supports multiple wallets of proof-of-work blockchains.

\subsection{Design of the Approach}
Our design operates on the principle of offloading computationally intensive header verification to a dedicated server.
This server generates succinct zero-knowledge proofs (specifically zk-SNARKs) attesting to the validity of batches of headers from a target blockchain (i.e., secondary blockchain).
These proofs are significantly smaller than the header data itself and are validated by a smart contract deployed on a primary blockchain with a smart contract platform.
Instead of directly verifying entire header chains, clients can rely on the primary blockchain's smart contract as a trusted intermediary.
By querying this smart contract and utilizing Merkle proofs, clients can efficiently verify transaction inclusion in the secondary blockchain, significantly reducing the resources needed for secure mobile wallet operation.

\subsection*{System Components}
\begin{figure}[t]
	\centering
	\includegraphics[width=\linewidth]{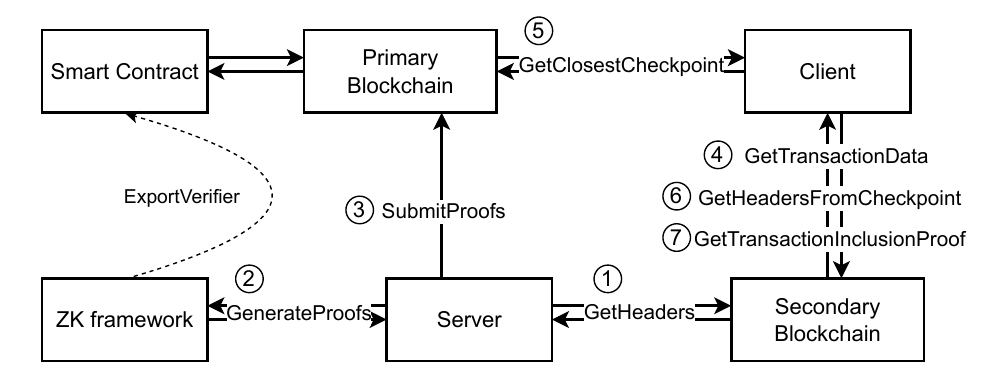}
	\caption{Overview of the components of our approach and how they interact.}
	\label{framework}
    \vspace{-1.5em}
\end{figure}

The system is composed of four key components (see \autoref{framework}) that interact between two types of blockchain as follows:
\begin{itemize}
    \item \textbf{\textit{Client}}: Represents the mobile wallet application.
    Without having full node capabilities, the client aims to efficiently verify transactions from a secondary blockchain.

    \item \textbf{\textit{Server}}: Acts as a proof generator.
    It monitors a secondary blockchain, downloads block headers, and generates zk-SNARK proofs of their validity in batches.

    \item \textbf{\textit{Smart Contract} (zk-SNARK Verifier)}: A smart contract deployed on a primary blockchain.
    This contract is responsible for verifying the zk-SNARK proofs submitted by the server and ensuring the correctness and continuity of the header chain batches.

    \item \textbf{\textit{ZK Framework}}: Represents the tools and libraries that generate zk-SNARK proofs.
    In our case, we utilize a Circom framework.
\end{itemize}

\medskip
Our architecture works two types of blockchains:
\begin{itemize}
    \item \textbf{\textit{Primary Blockchain}}: An EVM-compatible blockchain to process smart contracts.
    In our case, we utilize Ethereum.
    The primary blockchain hosts the smart contract verifier, acting as the trust anchor for the system.

    \item \textbf{\textit{Secondary Blockchains}}: These are blockchains whose transaction data needs to be verified.
    In our work, we consider proof-of-work blockchains and their required verifications of header chains; however, a similar approach could be employed for proof-of-stake blockchains.
\end{itemize}

\subsection{Interaction among Components}
The interactions of the components are as follows:
\begin{enumerate}
    \item \textbf{Header Retrieval}: The \textit{server} periodically retrieves batches of block headers from the \textit{secondary blockchain} (synchronization target).

    \item \textbf{Proof Generation}: Using a \textit{ZK framework}, the \textit{server} generates a zk-SNARK proof attesting to the validity of the retrieved header batch.

    \item \textbf{Proof Submission}: The \textit{server} submits generated zk-SNARK proofs to the \textit{smart contract} deployed on the \textit{primary blockchain}.
    This submission is a transaction on the primary blockchain and costs gas; hence we optimized our approach for cost efficiency.
    Also, note that before a proof submission occurs, the verifier smart contract is deployed on a primary blockchain, which is done by the \textit{server}.

    \item \textbf{Transaction Data Request}: When a \textit{client} wants to verify a transaction on the \textit{secondary blockchain}, it requests transaction data from the secondary blockchain.

    \item \textbf{Checkpoint Request}: The \textit{client} queries the primary blockchain via the \textit{smart contract} to get closest checkpoint.
    This checkpoint is the latest validated block header from the secondary blockchain, as verified and stored by the smart contract.

    \item \textbf{Header Retrieval from the Checkpoint}: Using the checkpoint header obtained from the \textit{smart contract}, the \textit{client} can efficiently retrieve a few subsequent headers from the secondary blockchains (if needed).
    This allows the client to build a partial header chain starting from a trusted checkpoint.
    
    \item \textbf{Transaction Inclusion Proof Retrieval and Validation}:
    The \textit{client} requests a Merkle inclusion proof for the target transaction from a full node of the secondary blockchain.
    The \textit{client} then uses this Merkle proof, along with the validated headers obtained in the previous step and the checkpoint header, to locally verify the transaction's inclusion in the secondary blockchain.
\end{enumerate}

These interactions enable clients to perform lightweight transaction verification.
The server handles the computationally intensive task of proof generation, and the primary blockchain provides a secure and transparent platform for proof validation and checkpoint storage.
The server focuses on the automated process of generating and submitting proofs for batches of secondary blockchain headers to the smart contract.
In contrast, the client focuses on the inclusion verification of a specific transaction by leveraging the validated header checkpoints stored on the primary blockchain and constructing a partial header chain.

\subsection{Header Chain Verification}
\begin{figure}[t]
    \centering
    \includegraphics[width=\linewidth]{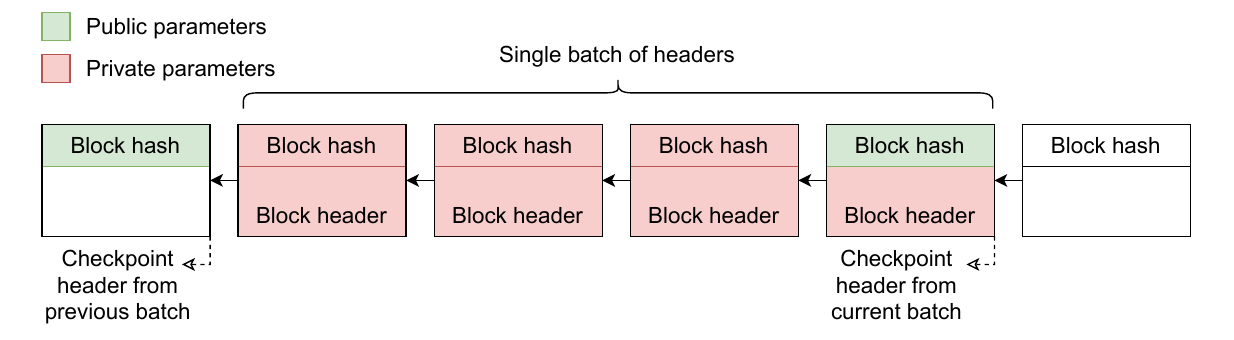}
    \caption{Header chain verification scheme for zk-SNARK prover and verifier.}
    \label{fig:header-verify}
    \vspace{-1.2em}
\end{figure}

The smart contract verifier is tasked with verifying the integrity of header chains.
All headers are private for one proof\footnote{Note that we do not leverage directly the zero-knowledge aspect of zk-SNARKs, but rather the property of succinctness, which ensures that the proof is smaller and verification time is faster compared to the original statement.} of the verifier (see \autoref{fig:header-verify}), only the hashes of the previous checkpoint header and the current checkpoint header are public.
The current checkpoint header serves to append it to this batch and for the overall continuity and integrity of the header chain as a checkpoint.
The verification process entails confirming that each header's previous block hash correctly points to the preceding block and that the target difficulty specified in the header exceeds its hash. 

\section{Implementation \& Evaluation}
\label{sections:eval}

\subsection{Implementation}
We made a proof-of-concept implementation, which comprises two main components, the server and the client.
The server generates proofs using a ZK framework.
We utilized \textbf{\textit{Circom/snarkjs}}~\cite{circom, snarkjs} with \textbf{\textit{Groth16}}~\cite{groth16} backend.
The server functionality also consists of querying the smart contract, for which we utilize web3.js.
The client is a mobile application that verifies the inclusion and synchronization of the targeted blockchain transactions.
The client application, designed for both iOS and Android, utilizes React Native and TypeScript. 
This choice enables cross-platform compatibility and a unified codebase, streamlining development.

\begin{table*}[ht]
    \centering
    \caption{Resource usage metrics for the various numbers of headers in a proof.}
    \label{tab:performance_metrics}
    \begin{tabularx}{\textwidth}{r|RR|RR|RR|RR|RR|RR|r}
        \toprule
        \textbf{Headers} & \multicolumn{2}{|c|}{\textbf{Compilation}} & \multicolumn{2}{|c|}{\textbf{Witness creation}} & \multicolumn{2}{|c|}{\textbf{Trusted setup}} & \multicolumn{2}{|c|}{\textbf{Ceremony contrib.}} & \multicolumn{2}{|c|}{\textbf{Proof generation}} & \multicolumn{2}{|c|}{\textbf{Proof verification}} & \textbf{Constraints} \\
        \midrule
        & Memory [\textit{MB}] & Time [\textit{s}] & Memory [\textit{MB}] & Time [\textit{s}] & Memory [\textit{MB}] & Time [\textit{s}] & Memory [\textit{MB}] & Time [\textit{s}] & Memory [\textit{MB}] & Time [\textit{s}] & Memory [\textit{MB}] & Time [\textit{s}] & \\
        \midrule
        2  & 1,167  & 6   & 12    & 0.051 & 2,010  & 17  & 413   & 7   & 2,773  & 3  & 312 & 0.373 & 179,270 \\
        4  & 2,289  & 12  & 14    & 0.093 & 3,239  & 29  & 435   & 13  & 4,555  & 4  & 314 & 0.357 & 358,540 \\
        8  & 4,521  & 23  & 149   & 0.185 & 5,039  & 55  & 498   & 24  & 6,590  & 7  & 316 & 0.365 & 717,080 \\
        16 & 8,922  & 47  & 408   & 0.382 & 8,814  & 115 & 976   & 49  & 10,613 & 15 & 316 & 0.349 & 1,434,160 \\
        32 & 17,699 & 98  & 810   & 0.771 & 15,006 & 236 & 1,338 & 96  & 20,684 & 30 & 320 & 0.350 & 2,868,320 \\
        64 & 35,312 & 204 & 1,614 & 1.447 & 29,783 & 567 & 1,619 & 196 & 38,552 & 58 & 321 & 0.384 & 5,736,640 \\
        \bottomrule
    \end{tabularx}
    \vspace{-1.5em}
\end{table*}

\subsection{Evaluation}
\label{section:eval-perf}
The evaluation focuses on assessing the framework's performance in terms of storage costs and computational overhead.
Benchmarks and comparative analysis with existing technologies demonstrate the advantages of the proposed system in real-world scenarios.
Our experiments focused primarily on the Circom/Snarkjs ZK framework and the smart contract components of our approach, as they are resource-intensive and crucial for functionality.
The client component performs minimal computation, so our client-side tests focused on assessing data requirements for synchronization across various checkpoint intervals.

\noindent
\textbf{\textit{Batch Submission Costs.}}
\autoref{fig:submission} shows the gas costs measured for different numbers of headers used for proof generation, submitted within a single transaction.
We conducted experiments with proofs containing 2,4,8,16,32, and 64 headers for the initial 5000 Bitcoin headers to determine the optimal transaction size, balancing the cost of initial synchronization and client synchronization expenses.
We can deduce that the number of headers covered by a proof does not affect the proof submission cost due to the zk-SNARK succinctness verification feature, which ensures that only two public parameters (two hashes) are produced in proofs regardless of the number of headers processed in batch.

During our experiments, we utilized the \texttt{--gas-report} functionality of the Foundry framework~\cite{foundry} (framework for developing smart contracts), which calculates the gas needed for transaction execution on the Ethereum network.
Our findings indicate that the cost of submitting multiple batches scales linearly with the number of batches involved.
The only limitation for the number of batches in a bundle is the maximum gas allowance of the Ethereum network per block, which according to our experimentation is $\sim$160 batches in a single transaction as of March 2025.

\begin{figure}[t]
    \centering
    \begin{tikzpicture}
        \begin{axis}[
            xlabel=Number of proofs in a single transaction,
            ylabel style={align=center}, ylabel=Price for proofs\\ submission (Gas),
            xmin=0, xmax=170,
            ymin=0, ymax=4e7,
            scaled ticks=false,
            xtick={0,16,32,...,170},
            ytick={0,5000000,..., 40000000},
            grid=both,
            grid style={line width=.1pt, draw=gray!10},
            major grid style={line width=.2pt, draw=gray!50},
            width=0.87\linewidth,
            height=0.5\linewidth
        ]
        \addplot[mark=,red,thick,dashed] plot coordinates {(0, 36e6) (176, 36e6)};
        \node[red, anchor=center] at (axis cs: 77, 32e6) {Current block gas limit};
        \addplot[mark=*,blue,thick] plot coordinates {
            (1, 260938)
            (16, 3638009)
            (32, 7240210)
            (48, 10842412)
            (64, 14444613)
            (80, 18046815)
            (96, 21649017)
            (112, 25251218)
            (128, 28853420)
            (144, 32455621)
            (160, 36057823)
            (176, 39660025)
        };
        \end{axis}
    \end{tikzpicture}
    \vspace{-0.5em}
    \caption{Price of submission for various number of proofs in transaction.}
    \label{fig:submission}
    \vspace{-1.2em}
\end{figure}
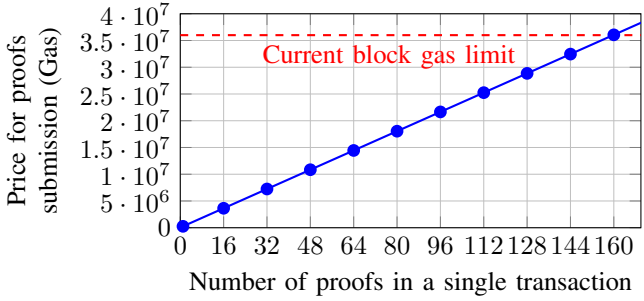

\begin{figure}[t]
    \centering
    \begin{tikzpicture}
        \begin{axis}[
            xlabel=Number of proofs in a single transaction,
            ylabel=Cost for monthly upkeep (Gas),
            xmin=0, xmax=70,
            ymin=1e7, ymax=7.5e7,
            scaled ticks=false,
            xtick={0,6,...,70},
            ytick={10000000,15000000,..., 75000000},
            legend style={at={(0.44,0.66)}, anchor=north west},
            grid=both,
            grid style={line width=.05pt, draw=gray!10},
            major grid style={line width=.2pt, draw=gray!50},
            width=0.9\linewidth,
            height=0.8\linewidth
        ]
        \addplot[mark=,red,thick,dashed,forget plot] plot coordinates {(0, 1.54e7) (70, 1.54e7)};
        \node[red, anchor=center] at (axis cs: 49, 1.27e7) {\small $\sim$ cost $1.54 \cdot 10^7$ [$\ddag$68 max]};
        \addplot[mark=*,red,mark size=1.2] coordinates {
            (1, 17878152) (2, 16653608) (3, 16245449) (4, 16040741) (5, 15918412)
            (6, 15836736) (7, 15778396) (8, 15734641) (9, 15700609) (10, 15673384)
            (11, 15651109) (12, 15632546) (13, 15616839) (14, 15603376) (15, 15591708)
            (16, 15581499) (17, 15572490) (18, 15564483) (19, 15557318) (20, 15550870)
            (21, 15545036) (22, 15539733) (23, 15534890) (24, 15530451) (25, 15526367)
            (26, 15522598) (27, 15519107) (28, 15515866) (29, 15512849) (30, 15510032)
            (31, 15507398) (32, 15504927) (33, 15502607) (34, 15500423) (35, 15498364)
            (36, 15496420) (37, 15494580) (38, 15492837) (39, 15491184) (40, 15489613)
            (41, 15488119) (42, 15486696) (43, 15485340) (44, 15484044) (45, 15482807)
            (46, 15481623) (47, 15480490) (48, 15479404) (49, 15478362) (50, 15477362)
            (51, 15476401) (52, 15475477) (53, 15474588) (54, 15473732) (55, 15472907)
            (56, 15472111) (57, 15471344) (58, 15470602) (59, 15469886) (60, 15469194)
            (61, 15468525) (62, 15467877) (63, 15467250) (64, 15466642) (65, 15466053)
            (66, 15465482) (67, 15464928) (68, 15464390)
        };

        \addplot[mark=,orange,thick,dashed,forget plot] plot coordinates {(0, 3.06e7) (70, 3.06e7)};
        \node[orange, anchor=center] at (axis cs: 48, 2.75e7) {\small $\sim$ cost $3.06 \cdot 10^7$  [$\ddag$135 max]};
        \addplot[mark=*,orange,mark size=1.2] coordinates {
            (1, 35492445) (2, 33060555) (3, 32250510) (4, 31845453) (5, 31602271)
            (6, 31440201) (7, 31324437) (8, 31237614) (9, 31170085) (10, 31116062)
            (11, 31071861) (12, 31035027) (13, 31003860) (14, 30977145) (15, 30953992)
            (16, 30933733) (17, 30915858) (18, 30899969) (19, 30885752) (20, 30872957)
            (21, 30861381) (22, 30850857) (23, 30841248) (24, 30832440) (25, 30824336)
            (26, 30816856) (27, 30809930) (28, 30803499) (29, 30797511) (30, 30791922)
            (31, 30786694) (32, 30781793) (33, 30777189) (34, 30772855) (35, 30768769)
            (36, 30764911) (37, 30761260) (38, 30757802) (39, 30754522) (40, 30751405)
            (41, 30748440) (42, 30745617) (43, 30742924) (44, 30740355) (45, 30737899)
            (46, 30735550) (47, 30733301) (48, 30731146) (49, 30729079) (50, 30727094)
            (51, 30725188) (52, 30723354) (53, 30721590) (54, 30719891) (55, 30718254)
            (56, 30716676) (57, 30715152) (58, 30713682) (59, 30712261) (60, 30710887)
            (61, 30709559) (62, 30708273) (63, 30707029) (64, 30705823) (65, 30704654)
            (66, 30703521) (67, 30702421) (68, 30701354) (69, 30700318) (70, 30699311)
        };

        \addplot[mark=,blue,thick,dashed,forget plot] plot coordinates {(0, 6.12e7) (70, 6.12e7)};
        \node[blue, anchor=center] at (axis cs: 48, 5.75e7) {\small $\sim$ cost $6.12 \cdot 10^7$  [$\ddag$270 max]};
        \addplot[mark=*,blue,mark size=1.2] coordinates {
            (1, 70978410) (2, 66119490) (3, 64498860) (4, 63688477) (5, 63202599)
            (6, 62878567) (7, 62647116) (8, 62473528) (9, 62338515) (10, 62230504)
            (11, 62142132) (12, 62068488) (13, 62006175) (14, 61952763) (15, 61906473)
            (16, 61865969) (17, 61830230) (18, 61798462) (19, 61770038) (20, 61744457)
            (21, 61721312) (22, 61700271) (23, 61681059) (24, 61663449) (25, 61647247)
            (26, 61632292) (27, 61618445) (28, 61605586) (29, 61593615) (30, 61582441)
            (31, 61571988) (32, 61562189) (33, 61552984) (34, 61544320) (35, 61536151)
            (36, 61528436) (37, 61521138) (38, 61514224) (39, 61507665) (40, 61501433)
            (41, 61495506) (42, 61489861) (43, 61484478) (44, 61479340) (45, 61474431)
            (46, 61469734) (47, 61465238) (48, 61460929) (49, 61456796) (50, 61452828)
            (51, 61449016) (52, 61445351) (53, 61441824) (54, 61438427) (55, 61435154)
            (56, 61431998) (57, 61428952) (58, 61426012) (59, 61423171) (60, 61420425)
            (61, 61417769) (62, 61415199) (63, 61412710) (64, 61410299) (65, 61407963)
            (66, 61405697) (67, 61403498) (68, 61401365) (69, 61399293) (70, 61397280)
        };

        \addlegendentry{64 header proofs}
        \addlegendentry{32 header proofs}
        \addlegendentry{16 header proofs}

        \end{axis}
    \end{tikzpicture}
    \vspace{-0.5em}
    \caption{Gas costs for monthly upkeep.}
    \label{fig:upkeep}
    \vspace{-1.3em}
\end{figure}
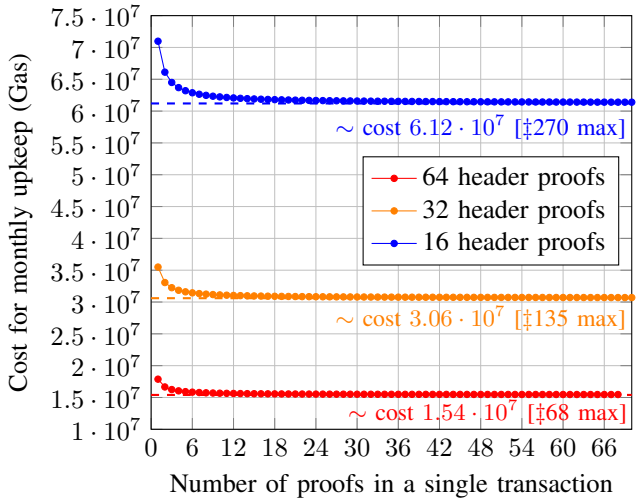

\noindent
\textbf{\textit{Price of Monthly Upkeep.}}
To keep the Bitcoin chain up to date, approximately 4320 headers must be verified monthly.
The gas cost of verifying a single proof is constant regardless of the number of headers the proof attests to.
As a result, doubling the headers per proof can approximately halve the monthly upkeep gas cost, as shown in \autoref{fig:upkeep}.
Given the overhead of Ethereum transactions, gas efficiency can be slightly improved by submitting multiple batches within a single transaction.
However, optimizing for lower gas costs by using larger batches and maximizing the number of headers processed by a batch in a single transaction comes at the expense of data freshness for wallet users.

\noindent
\textbf{\textit{Proof Creation.}}
Creating a single proof is a demanding process in terms of computational resources.
We monitored resource usage (see \autoref{tab:performance_metrics}), focusing separately on time and memory to compile the circuit (this is done only once for different numbers of headers in the proof) and the computation of the proof (for every set of headers as input for the proof).

Our tests showed that RAM requirements increase linearly (\autoref{fig:memory}) with the number of headers in each proof.
The tests carried out using the Circom/snarkjs ZK framework on an AMD Ryzen 9 7900X3D processor revealed (\autoref{fig:time}) that handling 64 headers for one proof takes 60\,s.

\begin{figure}[t]
    \centering
    \begin{minipage}{0.48\textwidth}
        \centering
        \begin{tikzpicture}
            \begin{axis}[
                xlabel=Headers,
                ylabel=Time (seconds),
                xmin=0, xmax=66,
                ymin=0, ymax=215,
                xtick={0,8,...,64},
                ytick={0,20,...,210},
                legend style={at={(0.02,0.96)}, anchor=north west},
                scaled ticks=false,
                grid=both,
                grid style={line width=.1pt, draw=gray!10},
                major grid style={line width=.2pt, draw=gray!50},
                width=0.95\linewidth,
                height=0.5\linewidth
                ]
                \addplot[smooth,mark=*,blue,thick] coordinates {
                    (2, 5.853)
                    (4, 11.313)
                    (8, 22.553)
                    (16, 46.600)
                    (32, 97.526)
                    (64, 203.524)
                };
                \addlegendentry{Compile}
                \addplot[smooth,mark=*,red,thick] coordinates {
                    (2, 2.362)
                    (4, 4.010)
                    (8, 7.327)
                    (16, 15.068)
                    (32, 29.511)
                    (64, 58.021)
                };
                \addlegendentry{Create proof}
                \end{axis}
        \end{tikzpicture}
        \vspace{-0.5em}
        \caption{Time metrics for various header counts within a single proof.}
        \label{fig:time}
    \end{minipage}\hfill
    \begin{minipage}{0.48\textwidth}
        \centering
        \begin{tikzpicture}
            \begin{axis}[
                xlabel=Headers,
                ylabel=Memory (GB),
                xmin=0, xmax=66,
                ymin=0, ymax=40,
                xtick={0,8,...,64},
                ytick={0,4,...,40},
                legend style={at={(0.02,0.96)}, anchor=north west},
                scaled ticks=false,
                grid=both,
                grid style={line width=.1pt, draw=gray!10},
                major grid style={line width=.2pt, draw=gray!50},
                width=0.95\linewidth,
                height=0.5\linewidth
                ]
                \addplot[smooth,mark=*,blue,thick] coordinates {
                    (2, 1.16729)
                    (4, 2.2894)
                    (8, 4.52156)
                    (16, 8.92211)
                    (32, 17.69969)
                    (64, 35.31234)
                };
                \addlegendentry{Compile}
                \addplot[smooth,mark=*,red,thick] coordinates {
                    (2, 2.7726)
                    (4, 4.55475)
                    (8, 6.58993)
                    (16, 10.61337)
                    (32, 20.68427)
                    (64, 38.55212)
                };
                \addlegendentry{Create proof}
                \end{axis}
        \end{tikzpicture}
        \vspace{-0.5em}
        \caption{RAM requirements for various header counts within a single proof.}
        \label{fig:memory}
        \vspace{-1.3em}
    \end{minipage}
\end{figure}

\noindent
\textbf{\textit{Storage Optimization.}}
Our framework offers significant storage optimization by replacing the need for multiple light clients on mobile devices.
In a conventional Bitcoin light client scenario (SPV), initial synchronization requires downloading, storing, and processing approximately 71\,MB per client (800,000 headers, each 80 bytes).
With current wallets that support multiple blockchains, this requirement can quickly escalate and become unmanageable.
Our solution leverages a single blockchain light node (i.e., server) as a reliable proof producer, with other blockchains dynamically synchronized as needed.
By only storing checkpoints, we can significantly decrease the data stored on the device.
In the case of one secondary blockchain (e.g., Bitcoin), the client only needs to store 3.5\,MB (which is 20 times less).

\section{Related Work}
\label{section:related}
The integration of zk-SNARKs into blockchain technologies has seen significant exploration aimed at enhancing security and efficiency.
Ben-Sasson et al.~\cite{zksnarks1} laid the foundational groundwork for zero-knowledge proofs within blockchain applications, allowing verification of transaction proofs without revealing underlying data.
Recent research has focused on mitigating the storage and computational overhead of blockchain clients.
Gudgeon et al.~\cite{blockchain-offloading} explored the use of side-chains for off-chain transaction handling, maintaining the security of the main blockchain.
Although conceptually related, our work differs in that we do not propose a separate chain; instead, we leverage a smart contract on the existing main chain for verification and checkpointing.

Bünz et al.~\cite{bulletproofs} introduced cryptographic accumulators (Bulletproofs) to reduce data and bandwidth requirements, addressing similar efficiency concerns as our mobile wallet framework.
Zk-relay~\cite{zk-relay} offers a lightweight, trustless cross-chain state-proving method using zk-SNARKs, relieving the target blockchain from the burden of verifying every block header.
This approach shares similarities with our approach in terms of off-chain proof generation, but we focus on
light mobile wallets and support of multiple PoW blockchains.

Plumo~\cite{plumo}, an ultra-light client based on zk-SNARKs validates state transitions over extended periods (approximately four months per proof).
This allows for rapid synchronization by verifying proofs instead of processing every block. However, Plumo, like many existing solutions (e.g., NIPoPoWs~\cite{light-clients}), is typically designed for a specific blockchain (Celo in the case of Plumo).
Moreover, Plumo long block ranges cause clients to manually download a lot of data since a recent checkpoint has occurred.
FlyClient~\cite{flyclient} is a non-interactive PoPoW, which overcomes the limitations of the superblock-based NIPoPoW protocol of Kiayias et al.~\cite{light-clients} and requires a PoW-specific blockchain design.
Similarly, projects like Mina~\cite{mina_protocol} (formerly known as Coda) leverage recursive zk-SNARKs to create a specialized constant-size blockchain via proofs, allowing highly efficient verification for clients.

\section{Discussion}
\label{sections:discussion}
\noindent
\textbf{\textit{Fork Resolution \& Smart Contract Security.}}
The smart contract verifier ensures security by selecting the strongest chain based on the cumulative work executed for header validation.
Although the attacker can submit valid batches of proofs (as an alternative chain), she would require more mining power than the main chain to win over stronger valid chains, which is, however, infeasible~\cite{zk-relay}.

\noindent
\textbf{\textit{Performance Overhead of Proof Generation.}}
As we showed in \autoref{section:eval-perf}, generating one proof of 64 headers takes around 1 minute with our setup, which significantly outpaces the blockchain's current block generation rate of 10 minutes on average.
This performance indicates that our setup can process proofs approximately 600 times faster than the Bitcoin block rate.
However, these results are currently limited by hardware resources, mainly by the memory limits.
Although further optimizations could enhance performance, initial synchronization, given that generating proofs for all Bitcoin blocks from the genesis up to block 887,640, would take approximately 10 days in our setup.

\noindent
\textbf{\textit{Parametrization.}}
The parameterization of the number of headers in a proof and the number of proofs within a transaction involves a trade-off between gas cost and finality time as observed by the wallet user.
Although proofs with more headers help reduce gas costs, they increase the computational load of clients caused by manual header validation between checkpoints and extend the time until the wallet confirms the most recent blocks.
For instance, 32 headers of Bitcoin correspond to around 5 hours, though this value ultimately depends on the particular secondary blockchain. 
However, scalability is constrained by the prover's RAM and CPU performance, which limits the upper bound of the number of headers in a proof.
Small but appreciable gas cost optimization can be gained by submitting multiple proofs in a single transaction.
Although small bundles offer some benefits, excessively large bundles yield diminishing returns in terms of gas cost reduction.
This optimization, again, comes at the expense of slower visibility of the most recent network updates.
Ultimately, the choice of these parameters comes down to the wallet user's preference for freshness, the prover's CPU and RAM constraints.

\noindent
\textbf{\textit{Costs for Maintaining Server.}}
As shown in \autoref{section:eval-perf}, publishing proofs and creating checkpoints incurs gas costs.
We estimate that monthly upkeep, considering appropriate checkpoint intervals and server computation costs, would require approximately 16,000,000 gas, which currently is 0.01 ETH ($\rightarrow$ 20 USD).
To cover these operational costs, users should pay a small subscription fee to provide the service.

\noindent
\textbf{\textit{Framework Extensions.}}
The framework is poised for expansion to accommodate even different consensus mechanisms than proof-of-work, which would require modification of the proof circuits.
This is planned in our future work.
Another promising development is the integration of parallel processing and automation for proof creation and submission.

\section{Conclusion}
\label{sections:conclusion}
In this work, we have designed and implemented an approach for lightweight mobile wallets, which uses zk-SNARKs and smart contracts to improve synchronization efficiency and reduce data demands compared to traditional multiple local light clients.
Our solution significantly reduces local data storage requirements by delegating the intensive task of header verification to a dedicated server and employing an Ethereum-based smart contract as a decentralized verification and checkpointing mechanism.
In the case of Bitcoin, our approach achieves a twenty-fold reduction in storage by maintaining only succinct checkpoints instead of whole header chains.

The evaluation demonstrated that the overall system scales efficiently, despite incurring computational demands associated with proof generation.
Batch submission of zk-SNARK proofs further contributes to gas cost optimization, providing a balanced trade-off between cost efficiency and the timeliness of network updates.
Future research may further optimize proof generation processes and extend support to a broader array of consensus mechanisms, such as proof-of-stake protocols.
These enhancements could amplify the applicability of our approach in diverse decentralized environments, ultimately contributing to the development of more secure, scalable, and resource-efficient mobile blockchain wallets and decentralized applications.

\bibliographystyle{IEEEtran}
\bibliography{IEEEabrv,bibliography}

\end{document}